\documentclass{camera}

\usepackage{graphicx}

\begin{document}

\title{Are there pulsars in the knee~?}

\footnote{E-mail address: erlykin@sci.lebedev.ru} 

\author{A.D.Erlykin~$^{(1,2)}$ \and A.W.Wolfendale$^{(2)}$}

\organization{$^{(1,2)}$ P.N.Lebedev Physical Institute, Moscow, Russia\\
$^{(2)}$ University of Durham, Durham, UK}

\maketitle

\begin{abstract}
Recent findings indicate that the Monogem Ring supernova remnant (~SNR~) 
and the associated pulsar 
PSR B0656+14 may be the `Single Source' responsible for the formation
of the sharp knee in the cosmic ray energy spectrum at $\sim$3PeV. We estimate the
contribution of the pulsar B0656+14 to the cosmic rays in the PeV
region and conclude that the pulsar cannot contribute more than 15\% to the cosmic ray
intensity at the knee. Therefore it cannot be the dominant source there and an SNR is 
still needed. 

We also examine the possibility of the pulsar giving
the peak of the extensive air shower (~EAS~) intensity observed from
the region inside the Monogem Ring. The estimates of the
gamma-ray flux produced by cosmic ray particles from this pulsar
indicate that it can be the source of the observed peak, if the
particles were confined within the SNR during a considerable fraction of its    
total age. 

We also estimate the contribution of Geminga and Vela pulsars to cosmic rays at the 
knee.
\end{abstract}
\section{Introduction}

A few years ago we suggested the `Single Source Model' to explain the remarkable 
sharpness of the knee in the cosmic ray energy spectrum at $\sim$3 PeV \cite{EW1,EW2}, 
a feature noticed even in the first publication on this subject, 46
years ago \cite{Kulik}. The model is based
on the assumption that a single, 
relatively recent and nearby supernova remnant contributes significantly 
to the cosmic ray intensity at PeV energies. The sharpness is due to the cutoff
in the energy spectrum of cosmic rays accelerated by SNR. Comparing the shape of the energy spectrum of cosmic rays from the 
Single Source and its total energy content with the model of SNR acceleration and 
the propagation of cosmic rays through the ISM \cite{EW9} we derived a likely interval of
distance (~230-350 pc~) and age (~84-100 kyear~) for the Single Source \cite{EW7}. 

On the basis of our estimates of distance and age we calculated the possible flux of 
high energy gamma rays from the Single Source and found that it is unlikely to be 
observed at sub-GeV and TeV gamma rays with gamma telescopes of the present sensitivity
\cite{EW7}. Among the sources which would satisfy these limits of distance and age we 
indicated the Monogem Ring and Loop I \cite{EW4}.  

Recently, Thorsett et al. \cite{Thors}, using the triangulation technique found the 
distance of the 
pulsar PSR 0656+14 associated with the SNR Monogem Ring. It is 288$\pm$30 pc and its 
spin-down age is $\sim$110 kyears, both of which are in remarkable agreement with our 
estimates for the 
Single Source \cite{EW7}. Thorsett et al. themselves claimed that the SNR Monogem Ring 
and its associated 
pulsar PSR 0656+14 can be the Single Source responsible for the
formation of the knee.

Armenian physicists have studied the sky near the Monogem Ring in the
sub-PeV and PeV range using the EAS technique and found a 6$\sigma$ excess
of the EAS intensity in one of their angular bins \cite{Chil1}. Since their bin
of $3^\circ \times 3^\circ$ is narrower than the size of the SNR it is
thought that the excess is not due to the extended source, but to
a discrete source, viz. the pulsar.
     
In this paper we analyse the possibility of the pulsar PSR B0656+14, 
associated with 
the SNR Monogem Ring, being the Single Source responsible for the knee
and also see to what extent the EAS excess (~the `Armenian peak'~) could come from the
same object.

\section{Contribution of an isolated pulsar to the knee}

At the beginning we consider the pulsar as an isolated neutron star.  
The energy spectrum of cosmic rays is calculated as
\begin{equation}
\frac{dN}{dE}=\int_0^T \frac{c}{4\pi} \frac{d^2N}{dtdE} S(t,T,E) \rho(t,T,E,R) dt
\end{equation}
Here {\em E} is the particle energy, $c$ is the speed of light, {\em t} is the time 
since the creation of the pulsar and $T$ is its spin-down age. 
$\frac{d^2N}{dtdE}$ is the particle emission rate, $S(t,T,E)$ is the 
survival probability against escape from the Galaxy and $\rho(t,T,R,E)$ is 
 the density of cosmic ray particles at a distance {\em R} from 
the source, emitted and observed at the time {\em t} and {\em T} respectively. 
Details of the calculations are given in \cite{EW12}.
Calculations have been made for the energy spectrum
of protons (~{\em Z} =1~) and oxygen nuclei (~{\em Z} =
8~), (~assuming that oxygen nuclei could, in fact, be taken from the
pulsar surface and accelerated by it~). The result is shown in Figure 1. 
\begin{figure}[htb]
\begin{center}
\includegraphics[width=7.5cm,height=7.5cm,angle=0]{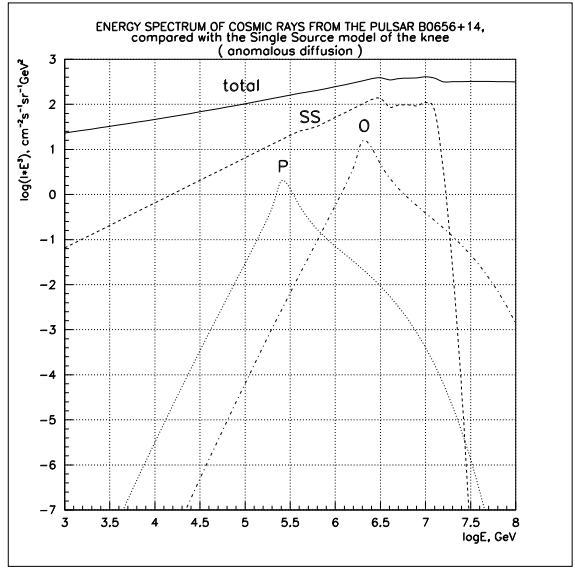}
\caption{\footnotesize The energy spectrum of cosmic rays from PSR B0656+14 observed 
at the present time and compared with the Single Source model of the
knee \cite{EW1,EW2}: full line - total energy spectrum of cosmic rays;
dashed line: energy spectrum of the Single Source, dotted line: energy
spectrum of protons accelerated by B0656+14, dash-dotted line - the
same spectrum for oxygen nuclei. The knee in the actual spectrum is at 
$logE(GeV) \approx 6.5$, close to our pulsar prediction for oxygen.}
\end{center}
\label{fig:puls1}
\end{figure}

It is remarkable that the spectrum of cosmic rays has a very 
sharp peak at a rigidity of 0.25 PV, i.e. a rigidity close to that of
the knee. It is the maximum rigidity of the particles 
emitted by the pulsar at the present time (~neglecting the time for
light to travel~).
The energy density contained in the proton and oxygen spectra is the same and equal to
$1.9\cdot 10^{-6} eVcm^{-3}$. Despite the fact that we 
normalized the energy transferred to the cosmic rays to the total loss of the rotation 
energy the cosmic ray energy density turns out to be small compared with 
the value of $\sim 2.24\cdot 10^{-4} eVcm^{-3}$ needed to form the knee in the SS model
\cite{EW7}. If, instead of energy density, we compare the intensity at the knee needed to ensure the observed 
cosmic ray flux and intensity in the peak for Z = 8, the difference becomes 
smaller, but it is still rather large (~$\simeq
7$~). We conclude, therefore, that 
the pulsar PSR 0656+14, if it is considered as an isolated neutron
star, can contribute up to $\sim15$\%
to the formation of the knee, due to the sharpness of its energy
spectrum and the 
closeness of its peak rigidity to the needed value of 0.4 PV, but it
seems not to be able to produce enough cosmic rays to be the dominant
source of the knee.

\section{The EAS intensity peak in the Monogem Ring region and the
possibility of associating it with the pulsar B0656+14}

\subsection{Observation of the peak}

Since the Monogem Ring SNR is located in our Local Superbubble, with
its low gas density,
 and it is not discrete, but an extended source, which occupies a substantial
part of the sky with an angular size of about 25$^\circ$, then we do not expect a 
measurable flux of high energy gamma quanta from it
\cite{EW7}. However, Armenian 
physicists looking for regions with an excessive flux of EAS at PeV
energies have found such a domain within the Monogem Ring SNR \cite{Chil1}
(~the `Armenian peak'~).
Their search bin had a size $3^\circ \times 3^\circ$, which is not point-like, but 
definitely smaller than the size of the Monogem Ring SNR itself. 
The magnitude of the excess was about 6 standard deviations and
therefore appears well founded statistically.  

The immediate idea, to be examined now, is that inside such an
extended object as the Monogem Ring SNR 
there is an additional discrete source of high energy cosmic rays,
which gives this excess. The most plausible discrete source within the SNR is the pulsar, specifically PSR B0656+14. Though the position of 
the peak is displaced from the present pulsar position, it is
reasonable to analyse the probability of this pulsar producing the 
observable peak. The inevitable diffusive scattering of
particles from the SNR and pulsar means that the Armenian peak must be due to
gamma rays or neutrons. 
     
It should be remarked that results of the experimental observation of the Monogem 
region is controversial. The armenian finding is the only positive result of a search,
supported marginally by the Moscow University group \cite{Zotov},
other people found nothing \cite{Alex,McKa,Wang,Ahar,Cui,Ant1,Ant2,Atki}. Therefore our  
estimates 
related to the Armenian peak are purely conventional, i.e. they have a meaning if the 
excess of EAS found by armenian physicists really exists.   
\subsection{Particles from an isolated pulsar}
From the data published in \cite{Chil1} we estimate a flux of particles giving rise to 
an observed EAS peak as $(1.2\pm0.4)\cdot 10^{-13} cm^{-2}s^{-1}$ \cite{EW12}.      
If these showers are produced by gamma quanta, their energy for the shower size 
$N_e > 10^6$ at 3200 m above sea level should be more than 1.07 PeV
\cite{Plyas}. If the showers are produced by neutrons, their energy
must be higher and for $N_e > 10^6$ at Aragats level should exceed 2.5
PeV \cite{Knapp}.

Since the pulsar B0656+14 is at a distance of about 300 pc from the
solar system, then if the observed gamma quanta or neutrons are
produced by protons from it, they can be born only 900 
years ago, i.e. at the present epoch. From the results presented in \S2 it
is clear that 
the pulsar B0656+14, if it is an isolated neutron star (~i.e. the
particles can diffuse freely from it~), cannot give particles 
above its peak energy of 0.25 PeV at the present epoch. Higher energy 
particles produced in the past would have the necessary higher energy
but they have already diffused for a long time and their density in the 
vicinity of the pulsar at the present epoch is very low. Heavier nuclei, if they are 
accelerated by this pulsar, and have higher total energy at the
present time, cannot help, since they have even smaller energy per
nucleon in the peak of their energy spectrum (~Figure 1~).

Therefore, if the pulsar B0656+14 is isolated it cannot give
the Armenian peak at energies above 1 PeV.   
\subsection{Particles from the pulsar associated with a SNR}   
There is a way to include into consideration higher energy particles born 
in the past, but which, however, produce gamma quanta or neutrons at
the present time and this is to 
associate the pulsar with a SNR, since they were both born in the same
SN explosion and reject the assumption that the pulsar can be regarded
as isolated. In this way we allow the produced particle to be trapped in the SNR
in the usual manner for SNR-accelerated particles
\cite{EW9,Koba2,Voelk}. The pulsar created as a result of the 
explosion is located within the shell close to the SNR morphological
center. We now assume that cosmic rays accelerated by the pulsar are also confined 
for the same time as those from the SNR. They are {\em all} released much later, begin to 
diffuse from internal regions of the SNR and eventually escape from the Galaxy. 
Their density in the vicinity of the pulsar is still high. In the process of diffusion 
through the ISM they produce qamma quanta which can be seen
now. This is the scenario.        

We calculate energy spectra of cosmic rays accelerated by the pulsar B0656+14
and observed now at the Earth assuming that the confinement time is the fraction
of the pulsar age. The results are shown in Figure 2.
\begin{figure}[htp]
\begin{center}
\includegraphics[width=12.6cm,height=12.6cm]{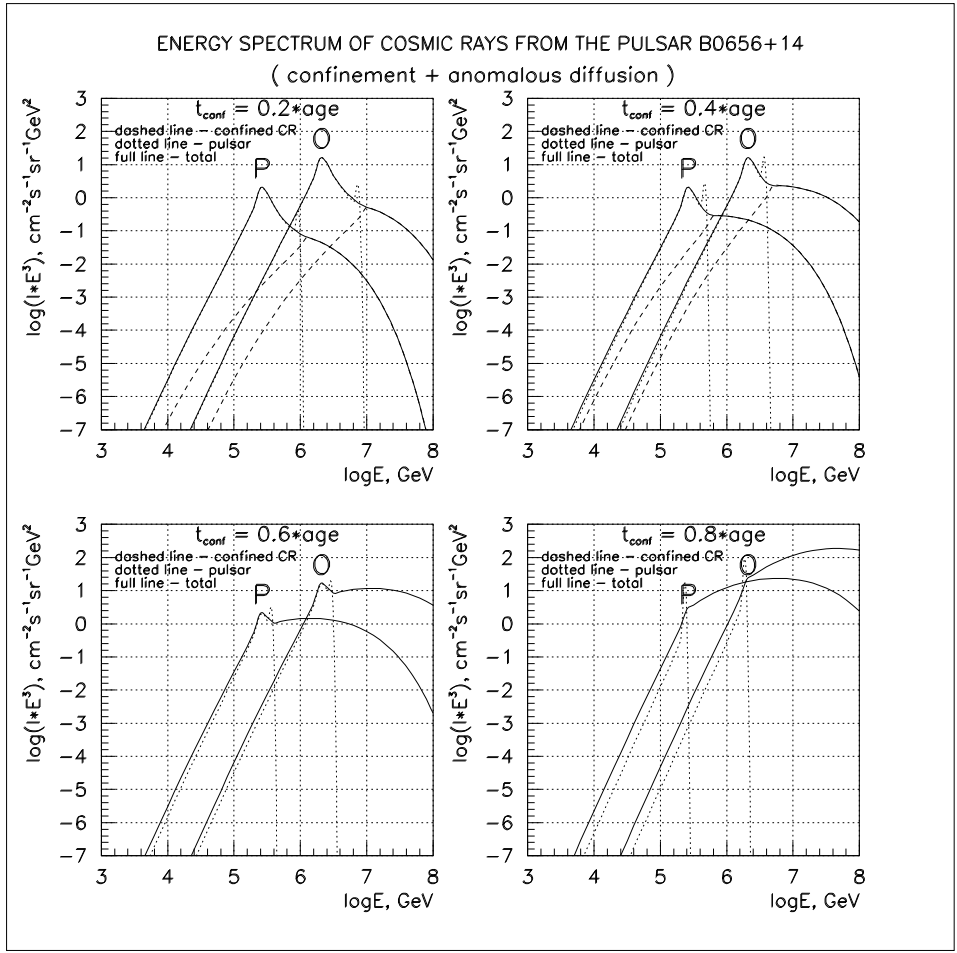}
\caption{\footnotesize The energy spectrum of cosmic rays from the
pulsar B0656+14 observed at the Earth, calculated for different
times $t_{conf}$, during which cosmic rays were confined within the
SNR shell and after that they were released, 
begin to diffuse through the ISM and escape from the Galaxy. Dashed lines indicate the 
contribution from the cosmic rays accumulated during the confinement time, dotted lines
show the contribution from the pulsar since the end of the confinement. Full lines show
the total spectrum composed of these two. The spectra are
 shown for the confinement time lasting (a) 0.2; (b) 0.4; (c) 0.6 and
(d) 0.8 times the pulsar age.}
\end{center}
\label{fig:puls2}
\end{figure}
It is seen that as the instant when cosmic rays are released
from confinement
 approaches the present moment, the higher 
energy cosmic rays remain within the region between the pulsar and the Earth and
can be the potential source of observed gamma quanta. Comparison with
 Figure 1 shows the `value' of the SNR trapping: particles above the
 knee (~at 3 PeV~) have now a higher intensity and thus this model is
 better able to explain the observation of particles of these high
 energies in practice.

\subsection{Gamma rays from the pulsar associated with the SNR}

We have calculated the expected
flux of gamma quanta with energy above 1 PeV as the integral along the
line of sight for gamma quanta produced in PP-collisions of protons
accelerated by the pulsar with the hydrogen atoms of the ISM:
\begin{equation}
F(>1PeV) = \int_0^R 2 c dr\int_{1PeV}^{E_{max}^0} \frac{dN}{dE}
\rho_{cr} \sigma_{in} n_{\gamma}(>1PeV) \rho_{ISM} dE 
\end{equation}     
Here $\frac{dN}{dE}$ is the energy spectrum of cosmic rays emitted by
the pulsar, 
confined during the time $t_{conf} = \Delta * age$, then released and, until the 
present time, 
survived after escape and diffusion (~Figure 2~). $\sigma_{in}$ is the inelastic 
cross-section of PP collisions, 
$n_{\gamma}(>1PeV)$ is the multiplicity
of gamma quanta with energy above 1 PeV, $\rho_{ISM}$ is the {\em mean} density of the 
target gas in
the ISM, taken as $3\cdot 10^{-3} cm^{-3}$ since B0656+14 is situated
in our Local Superbubble with its low gas density, $c$ is the speed of light. 

Calculation of the flux for $\Delta = 0.8$ gives a value of 
$3.7\cdot 10^{-15} cm^{-2}s^{-1}$,
which is less than the experimental value $\sim10^{-13}
cm^{-2}s^{-1}$ by more than an order of magnitude. However, the
calculations show a strong dependence of the gamma ray flux on the value
for the
time when the cosmic rays were released from confinement. For
instance, if $\Delta = 0.95$ the flux rises up to $2.0\cdot
10^{-13} cm^{-2}s^{-1}$ and exceeds the experimental value by the
factor of $\sim 2$. For  $\Delta = 0.99$ the flux is
 $2.2\cdot 10^{-12} cm^{-2}s^{-1}$, which is by an order of magnitude
higher than the previous value. 

We have also calculated the spectra of gamma quanta from interactions of nuclei, 
accelerated by the pulsar and
shown that they are about 2 times higher than that for protons.

Estimates show that electron inverse compton scattering on photons of the microwave 
background radiation or on X-ray quanta emitted by SNR and neutrons produced in charge
exchange $pp \rightarrow nX$ reactions or released from disintegrated cosmic ray nuclei
 cannot be the particles which give rise to the observed EAS peak \cite{EW12}.
The summary of our estimates are shown in Figure 3.    
\begin{figure}[htb]
\begin{center}
\includegraphics[width=7.5cm,height=7cm]{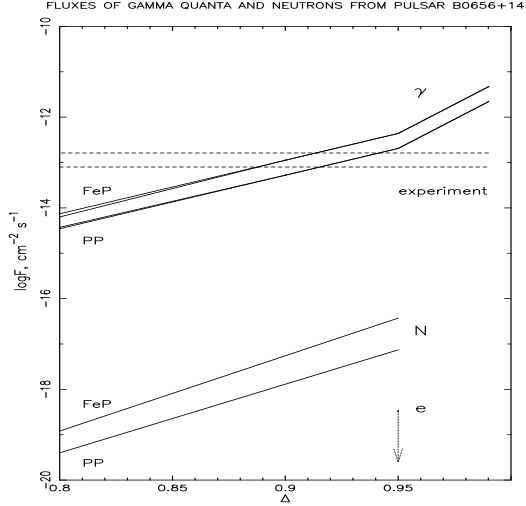}
\caption{\footnotesize Fluxes of gamma-quanta ($\gamma$) and neutrons ($N$) from PP and
 FeP collisions of protons and iron nuclei emitted by the pulsar B0656+14 as a function
 of $\Delta$ - 
a fraction of the total pulsar age durung which the emitted particles are confined
within a SNR. The inverse scattering of electrons ($e$) contributes much less 
(~shown by the dotted arrow~). The dashed lines indicate the experimental limits of the
flux found in \cite{Chil1} }
\end{center}
\label{fig:puls3}
\end{figure}

\section{Discussion}

We conclude that although the pulsar B0656+14 cannot be the domimant source
supplying cosmic rays in the knee region, it alone accelerates protons
which can produce gamma quanta observable as an excess EAS intensity
in the Armenian peak. Certainly cosmic rays from the associated SNR
Monogem Ring can contribute to this intensity. The only condition is
that cosmic rays from this pulsar should be confined by the associated
SNR during a considerable fraction of its age.
  
\subsection{Position of the Armenian peak and the pulsar B0656+14}

There is a potential difficulty in the association of the
Armenian peak with the pulsar. It is the mutual position of the region
where the EAS intensity peak is observed, the pulsar position and the
direction of its proper motion. Chilingarian et al. remark that
despite the fact that their intensity peak is inside the Monogem Ring
SNR it is displaced from the pulsar position by 8.5$^\circ$. It is rather 
far. Moreover, if the excess EAS intensity is due to
the interaction of cosmic rays produced by the pulsar in the past,
then the direction of its proper motion should be away from the
region where it was in the past. On the contrary, the direction of
B0656+14's proper motion is towards the region of the peak. 

However, even if everything is correct, our scenario with a long 
confinement could give an explanation
for the possible misalignment of the pulsar and the Armenian peak. 
The diffusion radius of PeV protons reached in
a few kyears is about 100 pc and a sphere of this radius can be seen from a  
distance of 300 pc at an angle of about 20$^\circ$. It means that PeV
cosmic rays have overcome the parent pulsar and its proper motion is
not necessarily connected with the regions of the highest cosmic ray
density. Within the sphere the density of cosmic rays is presumably 
uniform and any kind of local
ISM density perturbation or a local molecular cloud could create the
excess intensity of gamma quanta. 
\subsection{The connection between the pulsar and the SNR}
The confinement of high energy cosmic rays by the SNR raises other
intersting problems: how can the Monogem Ring SNR, which according to
our model accelerates cosmic rays only up to 0.4 PV rigidity, trap and
confine for a long time particles with rigidities up to $10^3$ PV ?
A possible idea for such a scenario is that SNR with their
large sizes and moderate magnetic fields are efficient in the
acceleration of relatively large fluxes of particles up to moderate
sub-PeV and PeV energies, whereas a pulsar which has much higher
magnetic fields in a much smaller volume, can accelerate smaller 
fluxes of particles reduced further by beaming, but up to much higher 
super-PeV energies. It means
again that the pulsar B0656+14 and, probably, other pulsars, might be
serious contenders for the sources of cosmic ray particles beyond the knee,
discussed often as the so called `second component' of cosmic rays in two-
or three-component models \cite{Wdowc,Ficht,Gaiss,Danil,Teran}.
\subsection{Possible contribution of other pulsars to the knee}
It is well known that some pulsars emit energetic gamma rays (~see \cite{Weeke} for a review~).
Bhadra examined the possibility for pulsars to form the knee and
concluded that Geminga and Vela are the most likely candidates
\cite{Bhadr}. We calculated energy spectra  of cosmic rays produced by
these pulsars, considering them as isolated pulsars; 
they are shown in Figure 4.

Since Geminga is older than
B0656+14 (~its age is $3\cdot 10^5$ y~), its maximum rigidity will be 0.19
PV, therefore it could contribute to the knee energies only if it
accelerates iron nuclei. In this case there is no room for the `second
knee' in the interval 10-20 PeV found by us and included in the
Single Source Model as an 'iron peak'.     
\begin{figure}[htb]
\begin{center}
\includegraphics[width=7.5cm,height=7.5cm,angle=0]{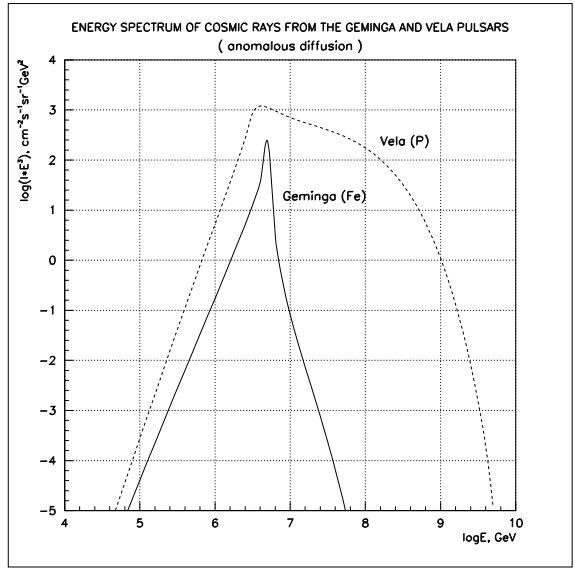}
\caption{\footnotesize Energy spectra of cosmic rays produced by
the Geminga and Vela pulsars. Full line: Geminga, accelerating iron nuclei,
dashed line: Vela, accelerating protons.}
\end{center}
\label{fig:puls4}
\end{figure}

Vela is much younger (~$\sim 10^4$ year~). Its maximum rigidity is
right in the knee region of $\sim$2.8 PV, so that it could contribute 
even by accelerating protons. The intensity can be adjusted
by introducing the efficiency of conversion of the pulsar rotation
energy to the accelerated particles at the level of 10\%, which is
a reasonable value. However, the Vela pulsar is associated with a young
SNR and most likely its accelerated  cosmic rays are still confined in the
SNR shell.  
\section{Conclusions}
We examined the possible contribution of several pulsars to the intensity of cosmic 
rays at the knee region. 
We conclude that the pulsar B0656+14 can contribute, but cannot be the dominant source 
in the knee responsible for its formation. 
Its contribution to the intensity of the Single Source, needed to form the sharp knee,
 does not exceed 15\%.  
The SNR associated with the Monogem Ring, rather than the pulsar, still remains the 
most likely Single Source 
which gives the dominant contribution to the formation of the cosmic ray energy 
spectrum in the vicinity of the knee.    

We have also examined the possibility of the pulsar B0656+14 giving
the peak of the EAS intensity, observed from
the region inside the Monogem Ring. The estimates of the
gamma-ray flux produced by cosmic ray protons from this pulsar
evidence that it can be the source of the observed peak, if the
protons were confined within the SNR during a considerable fraction
(~0.89-0.94~) of its total age. 

Other possible mechanisms for the production of particles which could
give rise to the Armenian peak were also
examined, but they were found improbable.

Geminga and Vela pulsars seem to be unlikely contributors at the knee.

\end{document}